\PassOptionsToPackage{table}{xcolor}
\documentclass[sigconf,nonacm,author]{acmart}

\expandafter\def\expandafter\UrlBreaks\expandafter{\UrlBreaks\do\/\do\*\do\-\do\~\do\'\do\"\do\-}
\usepackage{color}

\usepackage{wrapfig}
\usepackage{adjustbox}
\usepackage{multirow}
\usepackage{booktabs}

\usepackage{amssymb}
\usepackage{makecell}
\usepackage[table]{xcolor}
\usepackage{colortbl}

\newcolumntype{a}{>{\columncolor{blue!25}}c}

\begin{document}


\title{XaaS: Acceleration as a Service to Enable Productive High-Performance Cloud Computing}

\author{Torsten Hoefler}
\affiliation{%
  \institution{ETH Zurich \& Swiss National Supercomputing Centre (CSCS)}
  \country{Switzerland}
}

\author{Marcin Copik}
\affiliation{%
  \institution{ETH Zurich}
  \country{Switzerland}
}

\author{Pete Beckman}
\affiliation{%
  \institution{Argonne National Laboratory}
  \country{USA}
}

\author{Andrew Jones}
\affiliation{%
  \institution{Microsoft}
  \country{United Kingdom}
}

\author{Ian Foster}
\affiliation{%
  \institution{Argonne National Laboratory}
  \country{USA}
}

\author{Manish Parashar}
\affiliation{%
  \institution{Utah University}
  \country{USA}
}

\author{Daniel Reed}
\affiliation{%
  \institution{Utah University}
  \country{USA}
}

\author{Matthias Troyer}
\affiliation{%
  \institution{Microsoft}
  \country{USA}
}

\author{Thomas Schulthess}
\affiliation{%
  \institution{Swiss National Supercomputing Centre (CSCS)}
  \country{Switzerland}
}

\author{Dan Ernst}
\affiliation{%
  \institution{NVIDIA}
  \country{USA}
}

\author{Jack Dongarra}
\affiliation{%
  \institution{University of Tennessee}
  \country{USA}
}

\begin{abstract}
HPC and Cloud have evolved independently, specializing their innovations into performance or productivity.
Acceleration as a Service (XaaS) is a recipe to empower both fields with a shared execution platform that provides transparent access to computing resources, regardless of the underlying cloud or HPC service provider.
Bridging HPC and cloud advancements, XaaS presents a unified architecture built on performance-portable containers.
Our converged model concentrates on low-overhead, high-performance communication and computing, targeting resource-intensive workloads from climate simulations to machine learning.
XaaS lifts the restricted allocation model of Function-as-a-Service (FaaS), allowing users to benefit from the flexibility and efficient resource utilization of serverless while supporting long-running and performance-sensitive workloads from HPC.
%
%
\end{abstract}


\maketitle

\section{Introduction}

\setlength{\columnsep}{0pt}
\begin{wrapfigure}{r}{.25\linewidth} 
  \includegraphics[width=.25\columnwidth]{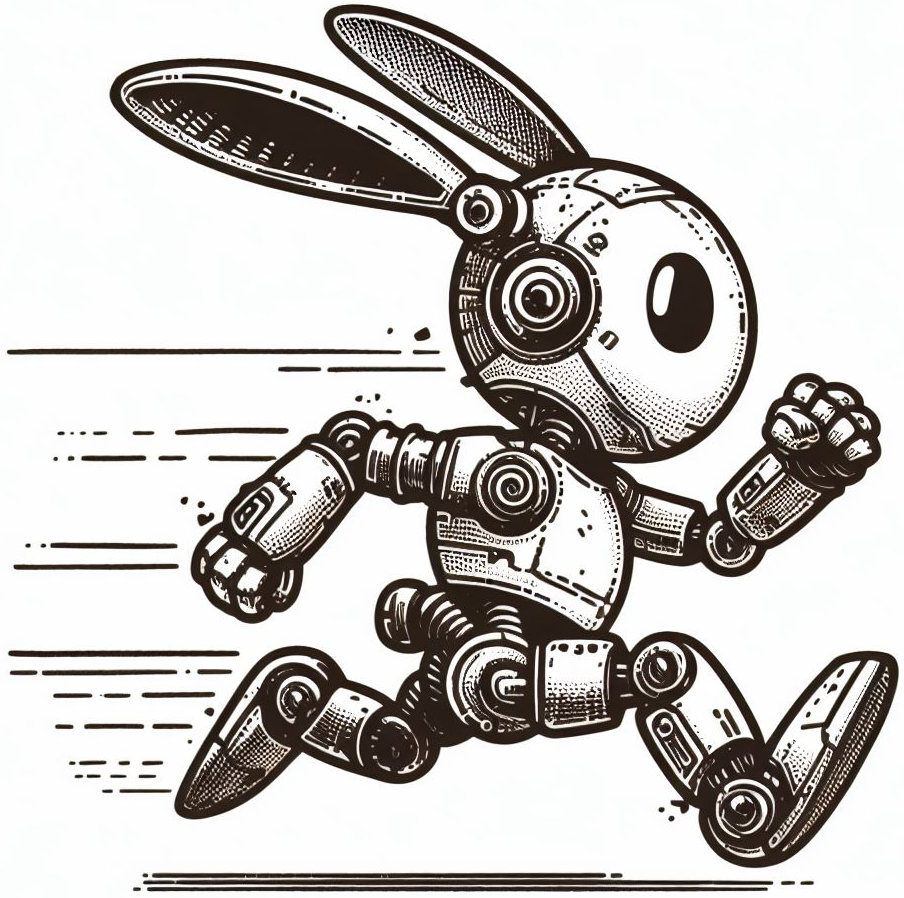}
\end{wrapfigure}
Acceleration as a Service (XaaS) is a recipe for enabling high-performance computing (HPC) workloads in the cloud. 
Cloud computing (“the Cloud”) provides the opportunity to offer computational capabilities as a simple transactional 
service, similar to how we use electricity or the internet.
Today's Cloud already offers a wide range of powerful services, from online storage to specific 
applications such as video calls or search. However, its performance is limited by inefficiencies in 
current Cloud architectures.
XaaS addresses those inefficiencies and enables the computation of high-performance accelerated workloads, ranging 
from simulations to AI/ML inference and training, as a high-performance cloud service capable of 
serving most demanding workloads.

\begin{figure*}
  \centerline{\includegraphics[width=0.95\textwidth]{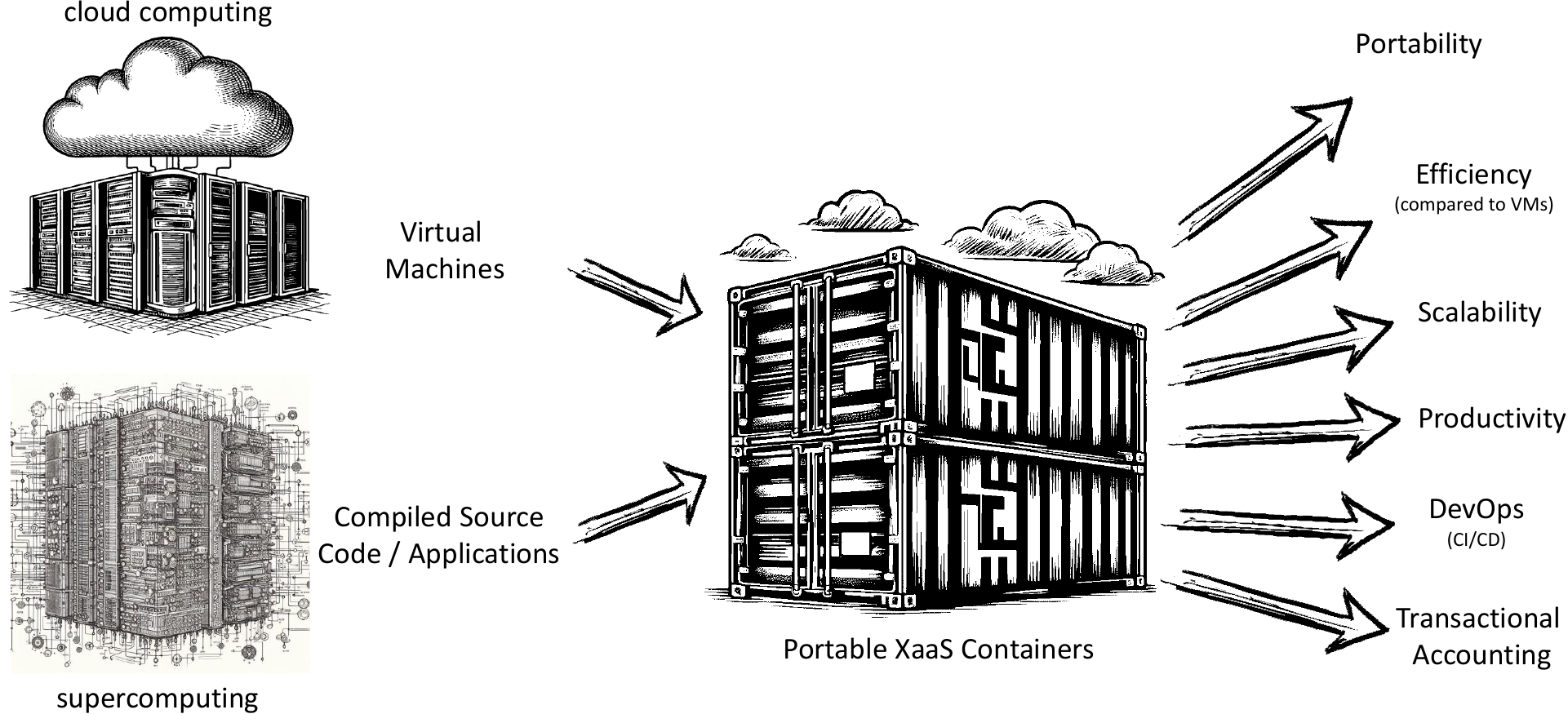}}
  \caption{Both Cloud and HPC converge to containers as an application and service deployment model.
  Containers bind all dependencies and system aspects (users, rights, etc.) into a single portable unit that can be flexibly deployed. XaaS enables HPC features for such containers.} 
  \label{fig:cloud_hpc_convergence}
\end{figure*}


%
XaaS provides different opportunities for people with different backgrounds and mindsets.
Members of the HPC community will find a vision for productive high-performance computing connecting today’s 
manually compiled-and-run HPC applications to a new world of automated high-performance containers running 
fine-grained transactional computations.
Members of the datacenter systems and cloud computing communities will find a vision for lifting standard 
container deployments seamlessly to low-overhead, high-performance accelerated infrastructures, 
enabling the fastest communication and specialized computing at the highest system utilization and 
reliability, whereby deployed containers utilize library interfaces and remote direct memory access (RDMA) 
technologies for specialized acceleration and communication with close-to-zero overheads compared to 
traditional bare-metal deployments.

Here, we define HPC workloads as resource-intensive and performance sensitive applications.
Traditionally, HPC systems were aimed at executing extremely demanding scientific computing workloads.
Recently, HPC systems have also been employed for data analytics, machine learning, and other workloads that, 
like scientific computing, require massive concurrency and rapid interprocess communication.
Supercomputing is the subset of HPC that uses the \emph{fastest and most powerful general purpose 
scientific computing systems available at any given time}~\cite{doi:10.1137/1.9780898719611}.
Cloud computing can be characterized by the desire to separate provider and user by a simple, clear, and automatable interface (ideally as simple as a power socket!) and by business and operations models designed to ensure that user requests can always be satisfied. To this end, cloud computing employs composable (micro)services that run in containers and interact through clearly defined interfaces (e.g., REST, JSON) that often however compromise performance. 

Applications that only rely on container and cloud service interfaces are called “cloud native.” Container creation, deployment, and management are largely handled by the de-facto standards Docker and Kubernetes. However, cloud service interfaces such as storage or machine learning inference are usually specific to the provider’s ecosystem.
%
Most modern cloud systems aim to offer an execution environment for cloud-native containers,
which is similar to an operating system’s interface to a process.
The Cloud Native Foundation seeks to define an interface in the spirit of the POSIX interfaces\footnote{\url{https://kubernetes.io/blog/2016/09/cloud-native-application-interfaces/}}.
This design is traditionally aimed at providing a productive ecosystem.
Only recently, performance has become a center of attention when using compute accelerators for 
demanding video processing tasks or AI/ML workloads.
Thus, the goals of modern cloud computing and HPC align well and could converge towards the 
same infrastructure.


\begin{table*}
\renewcommand{\arraystretch}{1.3} 
\centering
\vspace*{4pt}
\caption{Comparison point of existing Cloud offerings.}
\label{table}
\begin{adjustbox}{max width=\linewidth}
\begin{tabular}{|c|ccca|cc|@{}}
\hline
\multirow{2}{*}{\textbf{Cloud Overview}} & \multicolumn{4}{c|}{Generic} & \multicolumn{2}{c|}{Specialized}\\
\cline{2-5} \cline{6-7}
                    & \textbf{IaaS} & \textbf{PaaS} & \textbf{CaaS} & \textbf{FaaS} & \textbf{SaaS} & \textbf{DaaS}\\
\hline
Hardware Environment & \checkmark & \checkmark & \checkmark & \cellcolor{blue!25}{\checkmark} & \checkmark & \checkmark\\
Software Environment &  & \checkmark & \checkmark & \cellcolor{blue!25}{\checkmark} & \multicolumn{2}{c|}{\multirow{2}{*}{Pre-Configured Service}} \\
Bespoke Software     &  &  & \checkmark & \cellcolor{blue!25}{\checkmark} & \multicolumn{2}{c|}{} \\
Fine-grained Accounting &  &  &  & \cellcolor{blue!25}{\checkmark} & \checkmark & \checkmark\\
\hline
  \multirow{2}{*}{Example Services} & AWS EC2, Azure VMs & Google App & Azure AKS & \cellcolor{blue!25}{AWS Lambda, Azure} & Gmail& OneDrive, Box\\
                                  & GCP Compute Engine & Engine & AWS EKS & \cellcolor{blue!25}{Functions, GCP Functions} & Microsoft 365& Xignite for stock data\\
\hline
\end{tabular}
\end{adjustbox}
\label{tab:xaas}
\end{table*}

HPC and Cloud have progressed largely independently in the past.
Both according to their specialization: \textbf{The Cloud innovates in terms of business model, software packaging (containers), and productive ecosystems (e.g., cloud native) and HPC in terms of 
performance (e.g., RDMA) and abstractions for performance (performance libraries)}.
However, each field trails the other in other respects: for example, HPC has explored as-a-services 
abstractions~\cite{6280554} and is only just beginning to embrace the simpler deployment 
philosophy of containerized environments, while cloud started to explore ideas of RDMA.
Each feature was established in the other community a decade ago.
XaaS provides a way to accelerate this transition to a common architecture based on high-performance containers.
Figure~\ref{fig:cloud_hpc_convergence} shows a schematic overview of where each field is coming from and what containerized deployments could enable today or in the near future.
If those two communities do not join forces, they are bound to re-invent each other's methods.

All-in-all, the high-level architectural vision for a converged high-performance cloud with XaaS is 
based on three fundamental principles:

\begin{enumerate}
    \item \textbf{Performance portable containers (Infrastructure)}
    \item \textbf{High-performance communication and I/O (Input/Output)}
    \item \textbf{High-performance allocation, scheduling, and accounting (Invocation)}
\end{enumerate}

In the following, we outline three key techniques that can be used to build this architecture: 
Flexible hooked libraries and specialized builds can enable performance portability of the container 
infrastructure~\cite{10.1007/978-3-030-34356-9_5}.
RDMA and other direct memory access techniques can provide the lowest overhead interface to the outside 
world~\cite{10177421}.
Direct peer-to-peer allocations and high-performance scheduling and accounting can provide 
performant and available integration into a full system~\cite{10177421}.

\section{State of the Art}

We provide
detailed descriptions of HPC and CC, considering each field’s idiosyncrasies and commonalities. 

\textbf{HPC has traditionally supported demanding workloads in centralized datacenters.}
Supercomputers have long been used to serve the most demanding applications, such as 
weather prediction or the numerical simulation of complex structures; more recently, they are 
also used to train very large-scale AI models.
Due to the necessary large investment, supercomputers often pool the resources of many individuals at 
the regional or national level to address problems relevant to society.
While they are architected to run the largest jobs, they may spend much of their life running smaller
applications.
HPC centers have long led the design and development of large-scale systems, often in collaboration 
with system vendors.
HPC has driven the wide adoption of vector processing, massive parallelism based on commodity CPUs, 
general-purpose GPUs, and high-performance interconnects for multiple decades through long-term 
engagement with vendor partners. 

\textbf{Cloud emerged as a paradigm to sell compute cycles to a diverse set of customers},
ranging from anonymous customers with credit cards to long-term engagements.
The Cloud’s success in this endeavor has allowed it to realize, at scale, the vision of utility~\cite{5755602,armbrust2009above}
and grid computing~\cite{4738445} whereby computing as a service enables new services in 
many fields, including computational sciences~\cite{foster2017cloud}.
This approach changed the dynamics of IT businesses at large, giving startups a significantly lower barrier 
of entry compared to the dot-com days where the necessary CapEx proved to be a huge burden.
The Cloud’s aim to widen the customer base has led to a wide range of offerings at various 
levels of complexity and capability of compute services, encompassing Infrastructures (IaaS),
Platforms (PaaS), Containers (CaaS), and Functions (FaaS), as well as full application services 
such as Service Architectures/Software (SaaS) and Data (DaaS).
The focus is usually on reducing the barrier of entry and improving usability instead of performance, 
leading to relatively inefficient but simple web interfaces such as REST.
The latest push in this direction is the definition of cloud-native interfaces, for which performance 
and efficiency initially played only a secondary role.
Yet, due to economies of scale, cloud computing has become more performance-sensitive, 
especially in the emerging AI area.

\begin{figure*}
  \centerline{\includegraphics[width=0.95\textwidth]{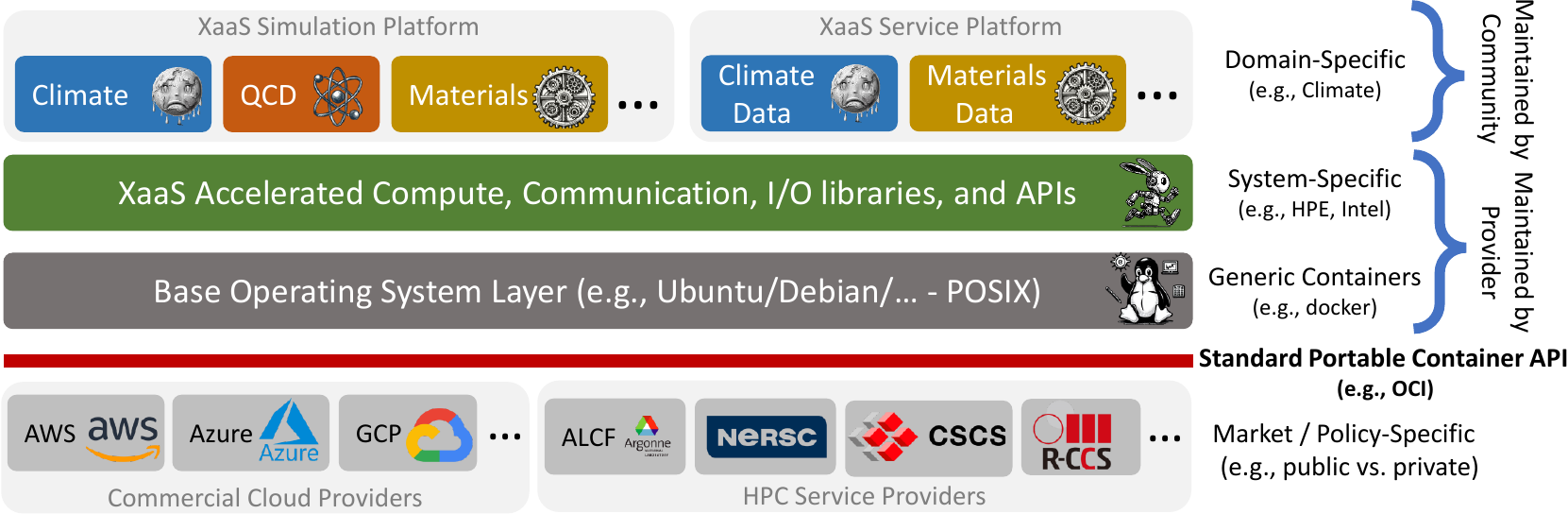}}
  \caption{XaaS ecosystem and components.}
  \label{fig:xaas_ecosystem}
\end{figure*}

It appears as if the market-driven Cloud field is moving organically towards a productive 
higher-performance environment in order to reduce costs of centralized services.
Meanwhile, many organizations in the community-driven HPC field remain in some sort of innovator’s 
dilemma whereby today’s traditional HPC environment, with its batch system setups, makefiles, 
and other venerable features, works well enough to throttle development and experimentation.
Yet this environment is increasingly not fit for purpose for emerging workloads that involve complex workflows or 
real-time computing.
Only a bottom-up movement, with potentially some top-down incentives, can change the field.
\emph{Only the right productive high-performance technology will move the community!}
\textbf{
We believe we need an architecture that enables portable, composable, and scalable workloads 
that allows users to build community-driven platforms at various levels.}
We believe that a fine-grained billable and containerized deployment, similar to FaaS (Table~\ref{tab:xaas}) but allowing much 
longer runtimes and large parallel jobs, would serve the community well.
While we do not prescribe implementations for such a service, we believe a microservices 
architecture could be used to implement and operate a XaaS infrastructure.
We continue by capturing and \textbf{contrasting the state of the art in both HPC and Cloud along multiple 
axes: usage, accounting, hardware, co-design, scheduling, and security}, and we outline a path to 
convergence towards productive high-performance accelerated cloud computing.

The basis of XaaS exosystem is a portable container API that abstracts interfaces from cloud and HPC providers together with a standard operating system layer (Figure~\ref{fig:xaas_ecosystem}).
The key addition in the XaaS software ecosystem is a system-specific set of accelerated APIs for compute (e.g., BLAS), communication (e.g., MPI, libfabric), and I/O (e.g., NetCDF) that are tuned to each target system and maintained by the provider. A recompilation layer would apply to the workflow of either building or deploying containers and is not shown. A standard XaaS layer enables portable accelerated domain-specific simulations and services in specific domains such as weather and climate, quantum chromodynamics and quantum simulations, or material sciences. Those domain-specific containers would be maintained by the respective community. 

On-premise HPC and Cloud are two extremes in a tradeoff between capital expense (CapEx) and operational expense (OpEx). A convergence of workloads and interfaces across both enables interesting opportunities to balance the two in the future.

\subsection{Containers}

Containers form an interesting design point in software deployment.
They emulate important parts of an operating system (e.g., a file system, processes, users) in a 
lightweight runtime that runs on a host operating system. 
The key is that containers provide a standardized clean and slim interface to the host OS and can thus be portable across many 
platforms and architectures.
They have their weaknesses, for example, excessive memory consumption due to limited sharing~\cite{2023upmbigdata}.
Yet, they form an interesting point solution in a complex design space.
Originally, computers ran individual applications that had to interface to all hardware directly. The emergence of multiprogramming in OSs then drove the adoption of portable interfaces (POSIX). The cloud started its journey by offering rented virtualized hardware as “Infrastructure as a Service” (IaaS), whereby customers would deploy their full OS as a virtual machine. Concurrently, HPC centers offered compute time to applications running in a machine-specific environment. Deploying a new application in such an environment routinely takes hours. The main difference to early Cloud VMs was that HPC applications were typically recompiled (to optimize them) for the specific machine and Cloud deployments were typically binary compatible (often x86). Such portable Cloud VMs are deployable in minutes on today’s cloud providers and provide the highest isolation as well as flexible choice of OS. When accounting for costs, VMs are typically charged by the hour.

Containers started as a way to package libraries and dependencies together with an application and quickly developed into an encapsulated OS-like environment for more complex services.
Most containers are significantly smaller than an OS VM (Megabytes vs. tens of Gigabytes) and can be deployed in seconds rather than minutes.
Their light weight enables a finer-grained accounting for “Container as a Service” (CaaS), often at a minute or second granularity.
Containers also support fast scale-up: a container image can be replicated to other machines to spin up more compute instances in seconds.
The latest development, \textbf{“Function as a Service” (FaaS) separates the deployment from the use of the containers}.
Requests simply invoke a function (which is defined in a container) that returns a result.
Initially, such functions were stateless, but they have recently been extended with local state and can of course access cloud APIs for persistent storage. The main benefit of FaaS is that the user is not involved in the deployment, which simplifies life for users while also enabling a fine-grained billing model on a millisecond scale for each function and allowing the operator to schedule function executions creatively. Such functions can be executed in containers or even micro-VMs for higher isolation. 

Portability requires that binary containers execute on different machines.
This needs both compatible container APIs as well as compatible binary executable formats.
Furthermore, many high-performance workloads require running parallel computations distributed to multiple machines; thus, containers need to be able to communicate efficiently over a network.
Today’s containers are based on standards defined by Docker and the Open Container Initiative (OCI) and portability is achieved by compiling containers for a given target architecture.
For example, the popular container repository DockerHub lists seven architectures ranging from x86 and ARM to IBM’s Z series. Yet, performance-critical workloads requiring lowest-overhead communications have received comparatively little attention so far.
We define \textbf{performance portable containers} as containers that achieve excellent performance on a variety of architectures. The term “excellent” admits various detailed definitions, such as “percent of peak” or “utilization”, which we purposefully leave open.

\subsection{Using resources}

\textbf{Users of HPC systems often engage with centers in a long-term (multi-year or decadal) relationship.}
This is partially due to what one could call “data gravity”, i.e., the hardship of moving massive amounts of data, but also due to the complexity of setting up a new environment and social aspects such as working practices and personal relationships.
HPC centers are interested in high utilization of their machines with economy-of-scale arguments.
Today’s batch systems typically reorder jobs to achieve the highest utilization.
This often leads to delays of large jobs but can also accelerate the scheduling of smaller jobs through “backfilling” a gap that a waiting larger job may cause.
This mode of operation is of course only amenable to run-to-completion jobs and cannot be used to operate online services or interactive sessions.
Important urgent or interactive applications such as disease and pandemic simulation~\cite{9651303}, real-time tsunami and earthquake simulation~\cite{9307622,YOSHIMOTO20121687},
and time-constrained data processing to guide future experiments~\cite{9307972,VESCOVI2022100606} require bespoke solutions on HPC systems or simply move to cloud systems.
A corollary of this mode of operation is that (1) system availability plays a secondary role -- HPC systems are often down for days at a time during working hours;
(2) even system reliability is secondary because checkpoint/restart is a viable mode of operation as long as the storage system is reliable.

\textbf{Cloud users want to remain flexible and able to change systems to allow the market to regulate pricing.}
Yet, cloud providers have little incentive to standardize their interfaces to achieve easy portability.
Proprietary interfaces and the high monetary costs of moving data out of the cloud result in some form of involuntary gravity towards a specific vendor.
Fundamentally, all cloud systems work similarly: they provide customers with a set of online services and microservices.
Even though providers can pass on their cost to customers if the market allows, optimizing the cost of foundational services can save millions of dollars in operational costs and is thus attractive to providers.
Sometimes, performance considerations even lead to service consolidation.
However, providers have only indirect incentives to improve the performance of customer workloads (who are charged per minute compute time).
In addition, new models such as FaaS allow providers to use their infrastructure more efficiently and thus lower costs while improving usability and enabling completely new service and billing models (pay as you go and scale down to zero).
For many cloud services, availability is critical and indeed non-negotiable as many important online systems run in cloud settings (e.g., credit card transactions, communication infrastructure).
Availability is achieved by highly resilient and redundant infrastructure that increases costs at all levels.
Scalability (aka “elasticity”) is also important, and while aggregate user demand may exhibit less variance than that of individual users, it ultimately requires costly idle resources at the time of each request.
In practice, resources are limited and requests can only be fulfilled if resources are available. 

\subsubsection{Opportunities of converged XaaS}
Both paradigms drive towards convergence: HPC requires increasingly reliable services, for example, for running performance- and availability-critical data systems such as the materials cloud\footnote{\url{https://www.materialscloud.org/}}, medical clouds\footnote{\url{https://www.cancergenomicscloud.org/}}, metagenomic analysis services~\cite{Keegan2016},
and earth virtualization engines~\cite{10301438}.
Cloud systems, on the other hand, already run batch jobs for background processing and are increasingly running large-scale bulk-synchronous AI training workloads in an HPC-like setting integrated with many services.
\textbf{One aspect to drive this convergence would be to consciously split workloads into interactive and non-interactive parts.}
For example, for a climate simulation, producing the data is a non-interactive component but analyzing and navigating the data is often driven by interactive discovery. 

Another important topic is ease-of-use.
Cloud architectures with containers in HPC would allow \textbf{communities to build their own platforms on top of a portable containerized environment.}
This way, \textbf{HPC providers could support high-performance container interfaces and communities could layer domain-specific services inside containers}, e.g., a climate simulation setup pre-installed in a container.
The \textbf{domain-specific layers could then drive scientific reproducibility and faster progress.}

\subsection{Scheduling}
By scheduling we mean the process of allocating resources to compute jobs. Ideally, a compute job would never wait for resources and always start immediately. Yet, having some jobs wait may greatly increase the average utilization in a compute system by shifting demand in time. From a user’s perspective, one needs to consider the whole response time. Humans operate in milliseconds and interactive requests, such as loading webpages, should return in that time-frame. Some high-performance jobs such as ML inference need to operate in those time frames while others, such as climate simulations, do not. Yet, even for non-interactive workloads, large amounts of computation must sometimes be provided in short time-frames, such as in emergency situations like natural disasters or pandemics.

\textbf{High-performance computers are usually used through batch systems that enable complex orchestration of scheduling} of different hardware types with advanced requests such as reservations.
They aim at high utilization and trade-off interactivity and waiting times and sometimes also performance (e.g., allocating arbitrary nodes instead of close nodes).
Some HPC centers are beginning to offer basic interactive services (at least a debug queue) and more and more are beginning to support “run forever” (server) type workloads.
Such workloads are often supported by bespoke solutions or cloud technologies in which some begin to employ a microservice architecture as infrastructure.
Thus, service workloads are slowly finding their way into HPC infrastructure and the job-based allocation model changes slowly.
One could say that a services allocation model requires committing some resources forever (at least very long periods of time) and not on a per-job basis. 

\textbf{Cloud initially scheduled only single VMs but moved quickly to groups of VMs to deploy microservices} with orchestrators such as Kubernetes.
The business was mostly focused around services for which reliability and availability is much more important than performance.
Thus, performance was often sacrificed for reliability, for example when allocating groups of VMs into different racks to avoid correlated failures in the power supply. In addition, service level agreements often state time expectations for submit-to-completion of interactive services. In addition, most cloud service providers run batch jobs to operate parts of the services that do not need to be interactive, for example, backups, nightly builds, or precomputed inference suggestions.

\subsubsection{Opportunities of converged XaaS}
\textbf{Portable high-performance containers would be beneficial for Cloud and HPC.}
For example, if one aims to collect low-rate sensor data over long periods of time, a normal cloud service is sufficient and cheap.
However, when it comes to processing or analyzing this data, a XaaS job is likely better. For serving results to external users XaaS may be appropriate if requests are computationally expensive, or a normal cloud service may be more cost-effective for data access requests. Cloud service providers also see growing demand for non-interactive and large compute jobs such as AI (re)training on incoming data. Both, HPC and Cloud providers need to analyze the requirements of interactive vs. non-interactive jobs carefully; XaaS could provide additional flexibility and new opportunities for both cases. 

\subsection{Accounting for resources}
\textbf{HPC resources are often provided by agencies} that make the acquisition of large resources easier than the money it would cost. Users propose research projects to acquire fixed allocations of resources to be consumed in a fixed time period. Those allocations cannot be repurposed for other things such as personnel. This approach, while it enables explorative high-risk research without the fiscal limitations, leads to a setup where research groups can acquire computational resources relatively cheaply (in terms of effort) but must invest their own people’s time into using them. Sometimes, users form consortia to support each other in such efforts, often focused on specific software (e.g., the US Lattice Quantum Chromodynamics or the Icosahedral Nonhydrostatic Weather and Climate Model Collaboration). However, such a setup often makes it hard to justify investing personnel resources into code optimizations, and performance consciousness thus varies largely across research groups and communities. Ultimately, users care about the total time and effort it takes to install, optimize, and execute their codes in a specific platform, rather than the aggregate efficiency of that platform.

\textbf{Cloud resources are acquired with real money paid by the users directly} in highly varying plans ranging from pay-as-you-go credit card transactions to year-long rentals for fixed provisioning. Many accounting schemes are complex and set up to have users spend more at a certain provider (e.g., free starting credits, through loyalty programs, or simply charging for outbound data copies while providing free inbound data copying). Performance has a direct pricing incentive and one can translate person-effort into money rather directly. 

\subsubsection{Opportunities of converged XaaS}
Funding agencies are already thinking about merging the two models. For example, NSF’s open science grid cloud and cloudbank operate with an allocation-based funding model at the user-facing side but buy the compute resources with real money from private and public clouds. This model exploits the flexibility to trade off CapEx and OpEx and the power of large-volume contracts. 

\subsection{Early hardware access, co-design, and code optimization}
\textbf{HPC centers often provide early access to hardware that they are going to deploy for users, in order to improve “application readiness”}.
Sometimes even pre-production hardware is offered in collaboration with vendors who are interested to optimize key applications for their architectures to provide the best price/performance ratio for compute centers and users. Achieving the best price/performance also drives system-level co-design that balances relatively high-level ratios, for example, network bandwidth per compute or storage bandwidth per compute. Engaging in more detailed hardware designs with vendors is complex because, despite the specialized purpose of HPC systems, their application mix is frequently diverse. Some HPC architectures were designed for specific applications (for example, IBM’s BlueGene was originally designed for biological applications), but they are generally used for a larger set of applications. Thus, co-design happens to a limited extent for some systems but is certainly not common practice today.

\textbf{Cloud service providers usually have access to early vendor designs and plans early on, but rarely expose that information to their users}.
One reason for this is that they want to run standard setups that they can scale quickly, cheaply, and at low risk to large user-bases. Diversity in special-purpose and early access hardware tends to hinder this scaling. Yet, today’s cloud providers are often first to roll out the newest technologies, and specialized compute may be supported if the user base is large enough or the service is profitable enough. For optimizing the workloads to the target architecture, cloud providers usually rely on HPC techniques such as libraries or compilers. Large markets (e.g., SQL databases or AI) can drive significant specialization and co-design at scale. Many of the architectures that providers use internally to provide such large-scale services are prime examples of co-design but are usually not exposed to the general public. 

\subsubsection{Opportunities of converged XaaS}
Hardware vendors pay more attention to larger markets and opportunities.
Thus, \textbf{an economy-of-scale argument benefits both HPC and Cloud}.
Today, most vendor attention is focused on cloud providers and thus growing the performance-awareness in this context would be beneficial for all.
\textbf{The opportunity to co-design hardware for AI and HPC workloads is huge and could be fueled by XaaS setups}, especially when combined with the flexibility of future chiplet-based architectures. 

\subsection{Security and Isolation}
\textbf{HPC systems traditionally do not focus on security and isolation} at the system level.
They either deploy unconnected (“air gapped”) systems or have relatively weak security standards because users are generally trusted after a careful admission check. HPC systems and users also generally trust system and network administrators. Yet, recently, with the increased importance of HPC in AI, health-care, and defense, security of HPC systems is receiving much more attention. 

\textbf{Cloud systems see security and isolation as mission-critical requirements}.
Encryption is often the default and trusted execution environments and even zero-trust environments are being rolled out. These capabilities are necessary because cloud providers (want to) admit anonymous users to their systems based solely on a credit card or other payment. Such users cannot be trusted. Furthermore, some big customers are not comfortable trusting the operator’s sysadmins. Thus cloud providers routinely implement special measures to implement “zero trust” settings (e.g., encrypt all stored data by default with user-provided securely handled keys). 

\subsubsection{Opportunities of converged XaaS}
As both HPC and Cloud have to deal with sensitive data and computations, \textbf{both will require performant security solutions}. Cloud systems could benefit from high-performance security systems to minimize overheads when providing privacy and isolation. 

\subsection{Summary}
From our state of the art discussion, we conclude that XaaS opens many opportunities when converging high-performance and cloud approaches and workloads.
XaaS \textbf{requires but also enables a culture change in the communities to enable layered high-performance software platforms driven by performance-portable XaaS containers.
Thus, we believe that we are at a perfect time and in a perfect setting to converge on XaaS architectures!}

\section{High-performance Accelerated Cloud Computing - A Road to Convergence}

\vspace{1pt}
\textbf{High-Performance Acceleration as a Service aims at a significant market, with AI as a service and HPC as a service being subsets, i.e., platforms that XaaS would enable. It will allow new solutions and scalable business on both the provider and user sides.}
Mainstream and most productive software development happens in the cloud space today and spawns a significant workforce that would provide value to the HPC community.
Yet, cloud development is often not aimed at the highest-performing solutions, which provides an opportunity for cloud computing to benefit from decreased cost and CO2 output for higher-performance solutions. 

Networking support and support for acceleration are two key areas of difference in HPC and cloud infrastructures today. While HPC has used both for decades, they are only now becoming relevant in generic cloud settings. XaaS should enable both in a manner that is consistent with the original vision of plug and play. 

We now refine the observations made above into three principal technical requirements that we already outlined in the introduction. We outline (a path toward) technical solutions for each of those requirements in the next section. 

\noindent\textbf{Performance portability} needs to ensure not only high-performing containers but also the ability to move containers between systems while still achieving good performance on all. These requirements imply low overheads for isolation and virtualization as well as support for native acceleration and specialization features at each system. 

\noindent\textbf{High-performance communication and I/O} is required to move data in and out of the portable environment. These capabilities are needed both for data stored in the provider’s system (e.g., storage access) and for data exchanged between different instances of containers during the computation (e.g., MPI communication). Low overhead schemes require operating system bypass solutions, well-known from HPC, whereby user-space applications directly communicate with the hardware. 

\noindent\textbf{High-performance allocation} systems are needed to reduce the waiting time for user applications. These allocation systems also need to support complex scheduling policies to differentiate interactive and batch jobs and potentially large requests that need to launch thousands of container instances into a large-scale job. Providing these capabilities will likely require decentralized or at least parallelized scheduling strategies.

\section{Enabling Technologies for XaaS}

We now briefly outline technologies and strategies that could be used to implement each of the three principal requirements (aka the “three Is”): \textbf{Infrastructure} for portable containers, fast communication and \textbf{Input/Output} to containers, and low-overhead and flexible \textbf{Invocation}.

\subsection{High-Performance Container Infrastructure and Input/Output}
Containers provide a simple and effective environment for software deployment by minimizing the interface to the outer (operating) system to clearly defined and slim calls. The Open Container Initiative (OCI) defines standards for container management. OCI offers hooks that allow dynamically linking system-specific libraries to containers during deployment. These  hooks enable the provider to bind system-optimized libraries to the container without the full system being  aware of the software running in the container. This additional layer of indirection allows programmers to extend the slim container interface with their own calls to performance libraries (e.g., BLAS or DNN). These capabilities can be enabled with Docker containers sitting on top of standard Linux namespaces and cgroups, for example. 

Existing HPC container infrastructures such as Apptainer (former Singularity) and Sarus are designed with performance in mind and take advantage of such features. Yet, there is no widely agreed standard for what libraries are supported for hooking across systems and what are the detailed interfaces and semantics of their calls. 

Having such a \textbf{flexible library hooks} interface also comes with some burdens. For example, not all libraries have the same hooks - if you want to hook into an MPI library, the interface will depend on whether the container binary was compiled against Open MPI or MPICH. Unfortunately, each has different ABI definitions. This problem can be solved, albeit at the cost of additional complexity, by implementing  multiple ABIs in each provider. The ecosystem could benefit from an ABI standardization. 

Library hooks solve the problem whenever performance-critical parts can be isolated into defined function calls. However, sometimes, complex application logic makes up for the majority of the time. In this case, compilers may be able to take advantage of specialties of the target architecture’s instruction set architecture (ISA). For example, NVIDIA’s H100 tensor cores offer much more functionality than V100 cores, and Intel CPUs that support AVX2 are more powerful than those that support AVX1 only. Using such features requires recompilation to the specific target architecture. Unfortunately, such recompilation is somewhat in conflict with the simple binary-deployment strategy of containers, and endangers the model of “compile and test on my laptop and then deploy on the largest supercomputers”.

One approach to consider for the ISA issue is \textbf{deployment recompilation}, similar to software deployment models in Gentoo Linux or Spack. One could attach a set of build scripts to each software to rebuild it at the target system using the system-specific optimizing compiler. This approach would greatly increase the complexity of container deployment in different execution environments - from simple binary ISAs and APIs to complex source codes. One could protect from failure by always including portable binaries that use only the lowest-common-denominator ISA features, but that approach would compromise performance portability. Another approach would be to ship precompiled source code in a compiler intermediate representation form (e.g., LLVM IR or DaCe SDFGs~\cite{10.1145/3295500.3356173}) that are then optimized at the target architecture. Other portability approaches such as WASM are probably not performant enough. 

\subsection{Fine-grained Invocation, Billing, Operations, and Integration}
Simple and fast invocation is key for accelerated high-performance cloud services. Such services often form workflows that are triggered or interfaced to from the outside. The connection to outside users could be offered through a web-service interface, for example based on a REST API such as FireCREST\footnote{\url{https://products.cscs.ch/firecrest/}}, which extends the established console interface with modern standard web services. Yet, REST must not be on the critical path due to its performance limitations, e.g., when transferring large data volumes. Yet, as a control interface, for example, to coordinate the deployment of a job or a virtual cluster, it is sufficient. Thus, as in Globus, the control channel may be REST, while the data channel employs high-performance protocols.

XaaS should support batch jobs as well as interactive services and enable deployment at various levels. While the typical deployment of XaaS may likely be FaaS, run-forever services could be deployed either at the IaaS, PaaS, CaaS, or FaaS levels. This variety of deployment levels will enable users to build and deploy their own high-performance microservice architectures in an environment that is most productive for them. Service providers can then support such executions or subsets of such executions (e.g., only FaaS) based on their business model.

\section{Opportunities}

We close by summarizing some of the opportunities of XaaS going forward. 
\textbf{A shared and compatible execution platform between cloud providers and high-performance computing centers provides many opportunities}.
It would widen the market and enable seamless access to various compute resources, independent of the provider.
Data location remains a challenging and somewhat fundamental issue, but decoupling the interfaces to data placement and to purchasing compute cycles will democratize big parts of the market.
Furthermore, \textbf{XaaS layers can enable scientific communities to distribute not only their source code but also their whole setup to others and thus enable seamless execution of their software across many architectures and providers at reasonable performance}.
A flexible scheduling and execution interface for XaaS maintains many of the benefits of FaaS workloads such that providers can increase their machine utilization; it will also enable large longer-running computations and sophisticated scheduling strategies.

\begin{acks}
\vspace{1pt}
The authors would like to thank Satoshi Matsuoka for valuable advice and comments.
\end{acks}

\bibliographystyle{ACM-Reference-Format}
\bibliography{references}

\section{Biographies}

\noindent\textbf{Torsten Hoefler} is a professor at ETH Zurich and the Chief Architect for Machine Learning at the Swiss National Supercomputing Center. His research interests revolve around high-performance artificial intelligence and computing systems. Hoefler received his highest degree in Computer Science from Indiana University. He is a fellow of the IEEE and ACM as well as a member of Academia Europaea. Contact him at \url{http://htor.ethz.ch/}. 

\noindent\textbf{Marcin Copik} is a senior PhD student at ETH Zurich. He received a Master’s degree from RWTH Aachen, Germany. His research interests are in high-performance solutions for serverless computing and cloud computing techniques for HPC. He received a Microsoft Research PhD Fellowship and ACM-IEEE CS George Michael HPC Fellowship. Contact him at \url{https://mcopik.github.io}.

\noindent\textbf{Pete Beckman} is Co-Director of the Northwestern University / Argonne Institute for Science and Engineering. His research interests include High-Performance System Software and Operating Systems. Beckman received his PhD in Computer Science from Indiana University. Contact him at \url{beckman@anl.gov}.

\noindent\textbf{Andrew Jones} is a Principal Program Manager at Microsoft at Redmond, WA, USA. His research interests include planning and delivery of large-scale high performance computing (HPC) services; technical and economic evaluation methods for HPC technologies and services; and the economic and human aspects of HPC, such as cost-value models and evolution of HPC skills. Jones received his BSc in Physics from the University of Manchester. He is a member at ACM SIGHPC. Contact him at \url{www.linkedin.com/in/andrewjones}.

\noindent\textbf{Ian Foster} is Distinguished Fellow and Director of the Data Science and Learning Division at Argonne National Laboratory in Lemont, Illinois 60439, USA, and Professor of Computer Science at the University of Chicago, Chicago, Illinois 60637, USA. His research interests include distributed and high-performance computing and their applications in the sciences. Foster received his PhD in Computer Science from Imperial College. He is a Fellow of the AAAS, ACM, BCS, and IEEE. Contact him at \url{foster@anl.gov}.

\noindent\textbf{Manish Parashar} is Director of the Scientific Computing and Imaging (SCI) Institute, Chair in Computational Science and Engineering, and Presidential Professor, Kalhert School of Computing at the University of Utah, Salt Lake City, UT, 84112. His research interests are in the broad areas of parallel and distributed computing and computational and data-enabled science and engineering. Parashar received his Ph.D. in Computer Engineering from Syracuse University. He is the founding chair of the IEEE Technical Community on High Performance Computing (TCHPC), and is Fellow of AAAS, ACM, and IEEE. Contact him at \url{http://manishparashar.org}.

\noindent\textbf{Daniel Reed} is a Presidential Professor and Professor of Computer Science and Electrical \& Computer Engineering at the University of Utah in Salt Lake City, Utah, 84117, USA.  His research interests include computational science, science and engineering policy, and high-performance computing.  Reed received his Ph.D. in computer science from Purdue University.  He is a fellow of the ACM, IEEE, and AAAS.  Contact him at \url{dan.reed@utah.edu}.

\noindent\textbf{Matthias Troyer} is Technical Fellow and Corporate Vice President at Microsoft Corporation in Redmond, WA. His interests include Quantum Computation, High-Performance Cloud Computing and AI acceleration for for science. Troyer received his PhD in Physics from ETH Zurich. Contact him at \url{matthias.troyer@microsoft.com}.

%

\noindent\textbf{Thomas Schulthess} is Director of the Swiss National Supercomputing Center (CSCS). His research interests include High-Performance and Cloud Computing. Schulthess received his PhD in Physics from ETH Zurich. Contact him at \url{schulthess@cscs.ch}.

\noindent\textbf{Daniel Ernst} is Director of System Architecture at Nvidia. His research interests include computer memory systems architecture, system performance modeling, and hardware/software co-design. Ernst received his PhD in Computer Science and Engineering from the University of Michigan. Contact him at \url{dane@nvidia.com}.

\noindent\textbf{Jack Dongarra} is Professor at the University of Tennessee Knoxville. His research interests include High-Performance Computing, Parallel Programming, and Numerical Algorithms. Dongarra received his PhD in Applied Mathematics from the University of New Mexico. He is a Fellow of the IEEE, ACM, SIAM, and AAAS, member of the US NAE, the US NAS, and a Fellow of the British Royal Society, as well as recipient of the ACM A.M. Turing Award. Contact him at \url{dongarra@icl.utk.edu}.

\end{document}